\def\thisversion{24 May 2021}
\documentclass[aps,prc,twocolumn,%
                nofootinbib,showpacs]{revtex4-2}

\usepackage{amsmath,amssymb}
\usepackage{newtxtext}
\usepackage{newtxmath}
\usepackage{graphicx}
\setcitestyle{citesep={,\!}}

\usepackage{xcolor}

\graphicspath{{../figs/}{figs/}}

\newcommand*{\hQ}{\hat{Q}}

\newcommand*{\pisub}{{\mspace{-1.5mu}\pi}}

\newcommand*{\Trans}{{\textsc{t}}}
\newcommand*{\Long}{{\textsc{l}}}

\allowdisplaybreaks

\begin{document}

\title{Gauge invariance of meson photo- and electroproduction currents revisited}

\author{Helmut~Haberzettl}
 \email{helmut.haberzettl@gwu.edu}
 \affiliation{Institute for Nuclear Studies and Department of Physics, The George Washington University, Washington, DC 20052,
 USA}

\date{\thisversion}

\begin{abstract}
An exact expression is derived for the meson production current off the
nucleon for real and virtual photons which cleanly separates, in a
model-independent manner, a tree-level expression that manifestly preserves
local gauge invariance in terms of a generalized Ward-Takahashi identity from
transverse final-state-interaction (FSI) terms. As discussed in some detail,
this exact formulation is particularly well suited for implementing
approximation schemes of exchange-current and FSI contributions (which in
practice generally will be necessary) at any level of sophistication without
violating local gauge invariance.
\end{abstract}



\maketitle


\section{Introduction}  \label{sec:introduction}

Local gauge invariance --- the requirement that physical observables are
invariant under local U(1) transformations of fields --- implies the very
existence of the electromagnetic gauge field $A^\mu$ and thus it is one of the
fundamental symmetries of electromagnetic physics~\cite{Peskin}. Conversely,
any violation of this symmetry in theoretical reaction models means that damage
has been done to the underlying description of the electromagnetic field. For
problems with complex reaction dynamics, in particular, where approximations
generally are unavoidable, local gauge invariance is oftentimes lost and some
effort must be made to repair this damage.

For pion photoproduction off the nucleon, the necessary and sufficient
condition for local gauge invariance was established by Kazes~\cite{Kazes1959}
in terms of a generalized Ward-Takahashi identity for the four-divergence of
the corresponding production current. This identity is easily verified for the
tree-level description of the process for structureless particles without form
factors, but it fails for more realistic descriptions with form factors. To our
knowledge, the earliest suggestion how to remedy the situation was given by
Drell and Lee~\cite{Drell1972} who proposed an additional nonsingular contact
current as a way to summarily account phenomenologically for all higher-order
contributions (like final-state interactions etc.) neglected at the tree level.
Gross and Riska~\cite{Gross1987} presented a repair scheme for gauge invariance
that reinterprets vertex dressing effects as self-energy contributions for
propagators connected to the vertex and modifies electromagnetic
single-particle currents such that this self-energy change is reflected in
their Ward-Takahashi identities. The \emph{ad hoc} Drell-Lee recipe was later
rediscovered independently by Ohta~\cite{Ohta1989} in a constructive procedure
implementing the electromagnetic interaction at the pion-nucleon vertex
utilizing expansions and making use of the minimal-substitution rule. In these
gauge-invariance-fixing schemes kinematic singularities are avoided for the
contact current by constructing suitable nonsingular $0/0$ expressions whose
four-divergences cancel gauge-invariance-violating terms. A seemingly different
method proposed by Nagorny and Dieperink~\cite{Nagorny1999} in terms of a
contact current written as a nonsingular interpolating integral can in fact be
shown to be functionally equivalent to the Drell-Lee-Ohta procedure. A
generalization of the latter approach was put foward by
Haberzettl~\cite{hh1997,hh1998} within a consistent field-theory framework for
describing electromagnetic meson production off baryons, including exchange
currents and hadronic final-state interactions (FSI), with mesons and baryons
as the relevant degrees of freedom. This framework also suggested a flexible
method for maintaining gauge invariance in a manifest manner, even when
approximating mechanisms like FSI or incorporating Regge
exchanges~\cite{hh2006,hh2011,Huang2012,hh2015}. The generalization also avoids
the `violation of scaling behavior'~\cite{Drell1972} of the original
Drell-Lee-Ohta recipe at high energies.

In all of these approaches the problem of maintaining gauge invariance and
approximating higher-order dynamical mechanisms are intricately linked. In the
present work, we will revisit the question of preserving local gauge invariance
even for problems with rich internal dynamics that generally require
approximations in practice by cleanly separating the issue of gauge invariance
from approximations of reaction dynamics. We will provide a novel formulation
of the production current that separates out, in a model-independent, exact
manner, manifestly gauge-invariance-preserving (GIP) tree-level expressions
from transverse final-state-interaction (FSI) terms. This formulation thus
allows devising efficient approximations to account for microscopic reaction
mechanisms like exchange-current contributions and final-state interactions
\emph{independent} of gauge invariance. Local gauge invariance
will always be preserved and never be at issue.

In Sec.~\ref{sec:ProdCurr}, we will briefly recapitulate expressions for the
production current relevant for the present approach. Section~\ref{sec:Gauge}
will then address local gauge invariance and provide the desired reformulation,
including a discussion of possible approximations that all maintain manifest
gauge invariance. The final Sec.~\ref{sec:Discussion} provides a summarizing
assessment.
%

%
\begin{figure}[t!]\centering
\includegraphics[width=\columnwidth,clip=]{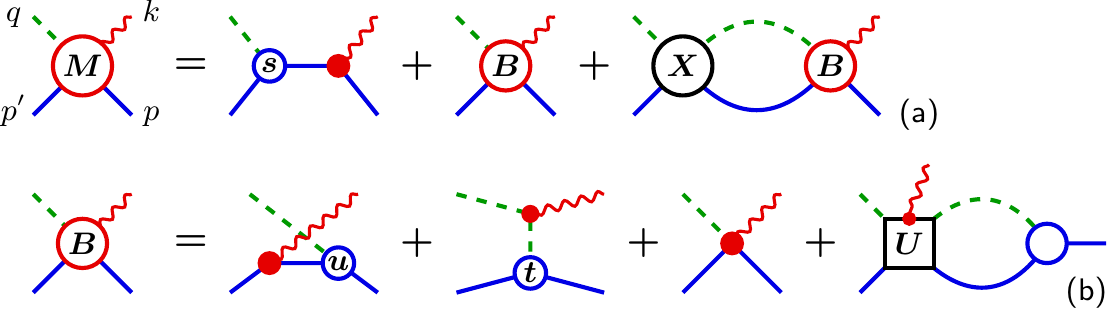}
  \caption{\label{fig:Mfig}%
  (a) Production current  $M^\mu$ of 
  Eq.~(\ref{eq:Mcoupled}). (b) Born-type mechanisms $B^\mu$ of
  Eq.~(\ref{eq:Bmu}). The last two diagrams provide the interaction current
  $M^\mu_c$ of Eq.~(\ref{eq:Mmuc}). Labels $s$, $u$, and $t$ refer to
  Mandelstam variables for the respective kinematic situations of hadronic
  vertices. Time proceeds from right to left to maintain one-to-one
  correspondence between equations and diagrams.}
\end{figure}
%

\section{Production Current}\label{sec:ProdCurr}

\begin{figure}[t!]\centering
\includegraphics[width=.64\columnwidth,clip=]{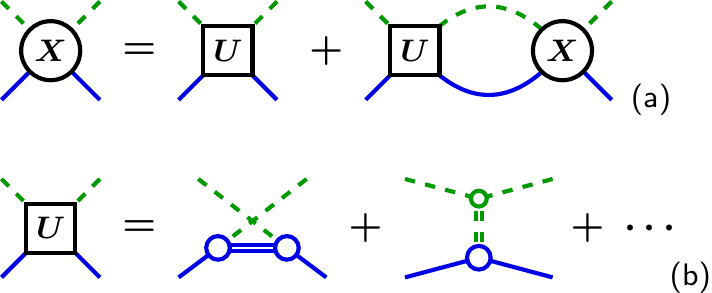}%
\caption{\label{fig:Xeq}%
 (a) Integral equation for the hadronic meson-baryon scattering amplitude $X$ of
 Eq.~(\ref{eq:Xamplitude}), with (b) nonpolar driving terms $U$; ellipses
 indicate higher-order loop contributions~\cite{hh1997}. Solid and dashed
 double lines subsume summations over all possible intermediate baryons  and
 mesons compatible with external meson-baryon states. }
\end{figure}

\begin{figure}[t!]
\centering
\includegraphics[width=.82\columnwidth,clip=]{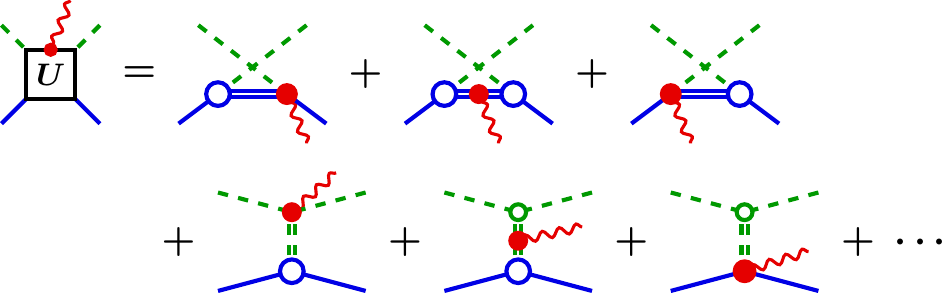}
\caption{\label{fig:UmuCurrent}%
Exchange currents $U^\mu$ obtained by coupling photons to the vertices and
propagators of the driving term $U$ of Fig.~\ref{fig:Xeq}(b).
}
\end{figure}

Taking nucleons ($N$) and pions ($\pi$) as
generic placeholders for baryons and mesons, the meson-production process
\begin{equation}
   \gamma(k) + N(p) \to \pi(q) + N(p')~,
  \label{eq:reaction}
\end{equation}
for real or virtual photons, is described by a production current
\begin{equation}
  M^\mu = M^\mu(q,p';k,p)
  \label{eq:Mmumomenta}
\end{equation}
that may be decomposed at the one-photon level as~\cite{hh1997}
\begin{equation}
  M^\mu = M_s^\mu + B^\mu + XG_0 B^\mu~,
  \label{eq:Mcoupled}
\end{equation}
represented by the diagrams of Fig.~\ref{fig:Mfig}(a). Here,  $M_s^\mu$ is the
$s$-channel production current and $B^\mu$ is the 2-particle-irreducible Born
mechanism,
\begin{equation}
  B^\mu = M_u^\mu+ M_t^\mu+ M_c^\mu~,
  \label{eq:Bmu}
\end{equation}%
depicted in Fig.~\ref{fig:Mfig}(b), that contains the usual $u$- and
$t$-channel contributions and the interaction current
\begin{equation}
  M^\mu_c = m^\mu_c + U^\mu G_0 F
  \label{eq:Mmuc}
\end{equation}
given by a Kroll-Ruderman-type elementary contact current $m^\mu_c$ (which may
be absent depending on the underlying hadronic coupling scheme) and a loop
integration over the pion-nucleon vertex $F$ and exchange currents $U^\mu$. The
ﬁnal-state interaction $X$ for meson-nucleon scattering is determined by the
integral equation
\begin{equation}
  X = U +U G_0 X~,
  \label{eq:Xamplitude}
\end{equation}
shown in Fig.~\ref{fig:Xeq}(a), driven by the 2-particle-irreducible exchanges
$U$ of Fig.~\ref{fig:Xeq}(b). (The exchange currents $U^\mu$ occurring in
Eq.~\ref{eq:Mmuc} are given by attaching a photon to this driving term $U$, as
depicted in Fig.~\ref{fig:UmuCurrent}.) The amplitude $X$ is not the full
physical meson-nucleon scattering amplitude $T$; the latter is obtained by
adding a fully dressed $s$-channel
pole term to $X$~\cite{hh1997}.\footnote{%
    In general, a splitting into pole and nonpole terms is not unique, however,
    consistency requirements of the field-theory approach of
    Ref.~\cite{hh1997} make it well defined here. Note in this context
    that while it is possible to write the FSI for the production current
    directly in terms of $T$, this will be at the expense of a very
    complicated, unusual
    $s$-channel contribution~\cite{hh1997,hh2011}. }
The intermediate free meson-nucleon Green's function is given by
\begin{equation}
  G_0 = t_\pisub(q_\pisub) \circ S(P-q_\pisub)~,
\end{equation}
where $t_\pisub$ describes the meson propagator with four-momentum $q_\pisub$
and $S$ provides the nucleon propagator depending on the remaining
four-momentum $P-q_\pisub$, where $P=k+p=q+p'$ is the total four-momentum of
the system. The symbol ``$\circ$'' stands for the convolution integral over the
loop momentum $q_\pisub$. Recalling that the terms pion and nucleon (and the
respective indices $\pi$ and $N$) are just representative placeholders for all
possible mesons and baryons, Eq.~(\ref{eq:Mcoupled}) is to be read as a system
of equations for all intermediate meson-baryon states compatible with the
external states, with all summation indices suppressed. As written, therefore,
the form of equations (\ref{eq:Mcoupled}) and (\ref{eq:Xamplitude}) is
deceptively simple to provide notational clarity. In reality, this is a highly
complex coupled-channel problem that generally cannot be solved exactly without
efficient approximation schemes.

%
%

%
%

\section{Local Gauge Invariance}\label{sec:Gauge}

Assuming that the description of the production current $M^\mu$ as given is
complete at the level of meson and baryon degrees of freedom, the necessary and
sufficient condition for its local gauge invariance is the generalized
Ward-Takahashi identity (gWTI) \cite{hh1997,Kazes1959},
\begin{equation}
  k_\mu M^\mu = - F_s S \hQ_N S^{-1} + S^{-1} \hQ_N S F_u  + t_\pisub^{-1}\hQ_\pisub t_\pisub F_t~,
  \label{eq:gWTI}
\end{equation}
where the functions $F_x$, for $x=s,u,t$, describe the pion-nucleon vertices
$F$ in $s$-, $u$-, and $t$-channel kinematics,
\begin{align}
  F_s &\equiv F(q,p';p+k)~,
  \nonumber\\
  F_u &\equiv F(q,p'-k;p)~,\\
  F_t&\equiv F(q-k,p';p)~,\nonumber
\end{align}
including hadronic coupling operators and isospin structures, with the order of
momentum arguments given as outgoing pion and nucleon momenta on the left of
the semicolon and incoming nucleon momentum on the right. In the notation here,
the operators with hats, $\hQ_N$ and $\hQ_\pisub$, indicate the charge
operators $Q_N$ and $Q_\pisub$ for the nucleon and pion, respectively, that
also inject the incoming photon four-momentum $k$ for the particle they pertain
to  into the equation at their respective locations. With external four-momenta
given as in Eq.~(\ref{eq:reaction}), this means that all momenta can be
recovered easily and unambiguously and do not need to be spelled out
explicitly, thus providing, for example,
\begin{align}
    F_s S \hQ_N S^{-1} &\equiv F(q,p';p+k) S(p+k)Q_N S^{-1}(p)~,
\end{align}
and similar expressions for $u$- and $t$-channel terms. In other words, the
charge operators $\hQ_n$ ($n=N,\pi$) as used here cannot be moved through the
expressions; they must remain at the locations where written. The inverse
propagators appearing in the gWTI (\ref{eq:gWTI}) imply that $k_\mu M^\mu=0$
for external on-shell hadrons thus providing current conservation. In general,
however, Eq.~(\ref{eq:gWTI}) holds true irrespective of whether hadrons are
on-shell and whether the photon is real or virtual.

Using the Ward-Takahashi identities~\cite{Ward1950,Takahashi1957} for the
electromagnetic single-particle currents of nucleons ($J^\mu_N$) and pions
($J^\mu_\pi$),
\begin{subequations}
\begin{align}
  k_\mu J^\mu_N &= S^{-1}\hQ_N- \hQ_N S^{-1}~,
  \\
  k_\mu J^\mu_\pisub &= t_\pisub^{-1}\hQ_\pisub- \hQ_\pisub t_\pisub^{-1}~,
\end{align}
\end{subequations}
and defining a longitudinal current
\begin{equation}
C^\mu_\Long
  =\frac{k^\mu}{k^2}\left(-F_s\hQ_N + \hQ_N F_u  + \hQ_\pisub F_t \right)~,
  \label{eq:CL1}
\end{equation}
the four-divergence of this current and of the $s$-, $u$-, and $t$-channel Born
terms  produces
\begin{align}
k_\mu &\left(M_s^\mu + M_u^\mu +M_t^\mu + C^\mu_\Long\right)
\nonumber\\
  &= - F_s S \hQ_N S^{-1} + S^{-1} \hQ_N S F_u  + t_\pisub^{-1}\hQ_\pisub t_\pisub F_t~,
  \label{eq:gWTIwithCL}
\end{align}
equal to the right-hand side of the gWTI (\ref{eq:gWTI}). Hence, rewriting the
production current (\ref{eq:Mcoupled}) equivalently as
\begin{align}
M^\mu
&=
\overbrace{M_s^\mu + M_u^\mu +M_t^\mu +C^\mu_\Long}^{\text{GIP}}
\nonumber\\
&\quad\mbox{}+ (M^\mu_c-C^\mu_\Long)+ XG_0 \big(M_u^\mu +M_t^\mu +M^\mu_c \big)~,
\label{eq:Mmu+C}
\end{align}
the terms labeled GIP already produce the gWTI. The remaining terms, therefore,
must vanish when taking their four-divergence. Formally,  we may then equate
the current $C^\mu_\Long$ of (\ref{eq:CL1}) with
\begin{equation}
C^\mu_\Long = \big[M_c^\mu +XG_0 \big(M_u^\mu +M_t^\mu +M^\mu_c \big)\big]_\Long~,
  \label{eq:CL2}
\end{equation}
where the bracket  $[\cdots]_\Long$ denotes the longitudinal part of the
enclosed current. The entire production current now reads equivalently
\begin{align}
  M^\mu &=  M_s^\mu+ M_u^\mu+ M_t^\mu+ C^\mu_\Long
  \nonumber\\
  &\quad\mbox{}
  + \big[M^\mu_c+ XG_0 \left( M_u^\mu+ M_t^\mu+ M^\mu_c\right)\big]_\Trans~,
  \label{eq:MmuexactCL}
\end{align}
where $[\cdots]_\Trans$ indicates that only transverse parts of the bracketed
current enter here, which do not contribute to the gWTI (\ref{eq:gWTI}).

\begin{figure}[t!]\centering
  \includegraphics[width=.95\columnwidth,clip=]{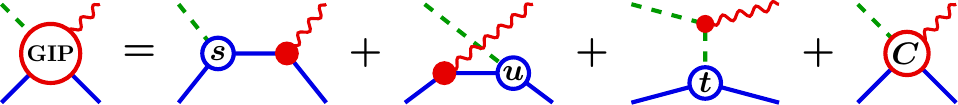}
  \caption{\label{fig:Mgip}%
  Gauge-invariance-preserving tree-level part of
  Eq.~(\ref{eq:MmuexactMC}), with the last diagram representing the contact-type
  GIP current $C^\mu$. These diagrams also correspond to the lowest-level
  approximation of the production current $M^\mu$ given in Eq.~(\ref{eq:Mmu0}). }
\end{figure}

For practical purposes and to circumvent the technical issues associated with
the $1/k^2$ singularity in $C^\mu_\Long$, we may rewrite
Eq.~(\ref{eq:MmuexactCL}) equivalently as
\begin{align}
  M^\mu &=  M_s^\mu+ M_u^\mu+ M_t^\mu+ C^\mu
  \nonumber\\
  &\quad\mbox{}
  + \big[M^\mu_c-C^\mu+ XG_0 \left( M_u^\mu+ M_t^\mu+ M^\mu_c\right)\big]_\Trans~,
  \label{eq:MmuexactMC}
\end{align}
where $C^\mu$ is the phenomenological, \emph{nonsingular}
gauge-invariance-preserving (GIP) current constructed in Refs.~\cite{hh1997,
hh2006} which reproduces the necessary condition
\begin{equation}
  k_\mu C^\mu = -F_s\hQ_N + \hQ_N F_u  + \hQ_\pisub F_t
  \label{eq:kCmu}
\end{equation}
to ensure the gWTI (\ref{eq:gWTI}). The particular details of the construction
in Refs.~\cite{hh1997, hh2006} are omitted here because they are not relevant
for
the present considerations.\footnote{%
      A concise description of the construction procedure for the GIP current
      $C^\mu$ is given in the Appendix of Ref.~\cite{hh2015}. In this context,
      note that the notation in Refs.~\cite{hh1997,hh2006,hh2015} is slightly
      different from the one employed here. Note in particular that $C^\mu$
      here is called $M^\mu_{\text{int}}$ in  Ref.~\cite[Appendix]{hh2015}, and
      that $C^\mu$ there is only the scalar coupling part of $C^\mu$ here.}
In fact, \emph{any} (contact-type, nonsingular) current $C^\mu$ that satisfies
(\ref{eq:kCmu}) will imply the \emph{manifest} preservation of local gauge
invariance. The gauge-invariance-preserving tree-level part of
Eq.~(\ref{eq:MmuexactMC}) is depicted in Fig.~\ref{fig:Mgip}. Because of the
subtraction of $C^\mu$ within $[\cdots]_\Trans$, the exact equation
(\ref{eq:MmuexactMC}) does not depend on the transverse part of $C^\mu$ which
is the only part that is model dependent. The subtraction thus avoids
double-counting of mechanisms already accounted for explicitly in $C^\mu$
outside of the $[\cdots]_\Trans$ bracket. This becomes particularly relevant
when one considers approximations of $M^\mu_c$ or final-state interactions ---
which generally will be necessary in practically applications. Some
straightforward approximation schemes are discussed in Sec.~\ref{sec:approx}.

%
%

Equation (\ref{eq:MmuexactMC}) is the main result of the present paper. By
construction, it provides an exact expression for the production current that
is manifestly gauge invariant. Most importantly for practical purposes, the
gauge invariance will remain true for \emph{any} approximation within the
transverse brackets $[\cdots]_\Trans$. The construction, in particular, shows
that explicit final-state interactions occur here only in manifestly transverse
contributions. This is similar to what was suggested by the approximation
scheme of Ref.~\cite{hh2006}. However, the present result shows that this
property is generally true, independent of any approximation because of the
$C^\mu$ subtraction in $[\cdots]_\Trans$. Implicit longitudinal FSI
contributions are subsumed in $C^\mu_\Long$ of Eq.~(\ref{eq:CL1}) via its
formal equivalence with Eq.~(\ref{eq:CL2}).

Physical on-shell matrix elements,
\begin{equation}
  \mathscr{M} = \epsilon_\mu M^\mu~,
\end{equation}
are formed with transverse states $\epsilon_\mu$, either as the transverse
photon polarization four-vector for real photons or, in the case of
electroproduction, as the Dirac current of the electron between on-shell
spinors times the photon propagator, $\epsilon^\mu=\bar{u}(p'_{\!e}) \gamma^\mu
u(p_{\!e})/k^2$, where $k=p_{\!e}-p'_{\!e}$ is the momentum difference
transferred from the incoming and outgoing electron momenta to the photon. In
both cases, the transversality condition $k_\mu \epsilon^\mu =0$ holds true and
thus only transverse parts of the production current contribute for physical
matrix elements,
\begin{align}
  \epsilon_\mu M^\mu &=\epsilon_\mu\big[ M^\mu_s +
  M_u^\mu+ M_t^\mu+ M^\mu_c
  \nonumber\\
  &\qquad\qquad\mbox{}+
   XG_0 \left(M_u^\mu+ M_t^\mu+ M^\mu_c \right)\big]_\Trans~,
   \label{eq:Transpol}
\end{align}
irrespective of whether the photon is real or virtual.

For the physical amplitude $\mathscr{M}$, therefore, the fact that the
gWTI~(\ref{eq:gWTI}) was explicitly enabled via the tree-level contribution of
Eq.~(\ref{eq:MmuexactMC}) by adding and subtracting the transverse parts of the
GIP current $C^\mu$ is quite irrelevant here since the current in the
transverse brackets $[\cdots]_\Trans$ of (\ref{eq:Transpol}) is still the full
production current $M^\mu$ stipulated to provide the gWTI. The situation will
change, however, if any of the mechanisms of $[\cdots]_\Trans$ in the exact
equation (\ref{eq:MmuexactMC}) is approximated --- which in practice generally
will be unavoidable because of the complexity of the contributing mechanisms.

\subsection{Approximations}\label{sec:approx}

By construction, any approximation of the transverse current $[\cdots]_\Trans$
in (\ref{eq:MmuexactMC}) will preserve full local gauge invariance. In general,
however, such approximations will make the dependence on the GIP current
$C^\mu$ manifest, in contrast to the exact amplitude (\ref{eq:Transpol}).

The simplest approximation in (\ref{eq:MmuexactMC}) is to drop the entire
$[\cdots]_\Trans$ current and replace the full current by the tree-level
expression
\begin{equation}
M^\mu =  M_s^\mu+ M_u^\mu+ M_t^\mu+ C^\mu
  \label{eq:Mmu0}
\end{equation}
depicted in Fig.~\ref{fig:Mgip}. This approximation thus complies with the
generic topological structure of a single-meson production
current~\cite{GellMann1954} and it provides the minimal structure necessary for
producing the gWTI~(\ref{eq:gWTI}). Effectively, this corresponds to the
replacement $ M^\mu_c + XG_0 B^\mu \to C^\mu$, with the contact-type GIP
current $C^\mu$ providing here a phenomenological description for all
exchange-current and FSI contributions depicted in Fig.~\ref{fig:Mfig}.  As the
description in the Appendix of Ref.~\cite{hh2015} shows, if necessary, $C^\mu$
will also account for a Kroll-Ruderman-type elementary contact current, as
required, for example, for pseudovector pion-nucleon coupling in pion
production. Moreover, additional phenomenological transverse contributions
$T^\mu$ may be added without affecting gauge invariance, $C^\mu \to C^\mu +
T^\mu$, to better model the problem at hand~\cite{hh2006}. Examples of
applications of this tree-level approximation are given in
Refs.~\cite{hh1998,Nakayama2004,Nakayama2006,Oh2008,Huang2013,Kim2014,Wang2016,Wang2017,Wei2019}.

A much more sophisticated approximation is to let $M^\mu_c\to C^\mu$, resulting
in
\begin{align}
  M^\mu &=M_s^\mu+ M_u^\mu+ M_t^\mu+ C^\mu
  \nonumber\\
  &\mbox{}\qquad
  +\big[XG_0(M_u^\mu+ M_t^\mu+ C^\mu)\big]_\Trans~,
  \label{eq:Mmu1}
\end{align}
where the dynamical details of the exchange-current contributions $U^\mu G_0 F$
depicted in Fig.~\ref{fig:Mfig}(b) are frozen out and summarily accounted for
by the phenomenological current $C^\mu$. However, hadronic final-state
interactions are fully incorporated. A variant of this approach was used for
pion photoproduction in Ref.~\cite{Huang2012}.

At a still more sophisticated level, one needs to explicitly consider the
exchange currents $U^\mu$ contained in $M^\mu_c$ of Eq.~(\ref{eq:Mmuc}), as
depicted in Figs.~\ref{fig:Mfig}(b) and (\ref{fig:UmuCurrent}). Structurally
this results in
\begin{align}
  M^\mu &=  M_s^\mu+ M_u^\mu+ M_t^\mu+ C^\mu
  \nonumber\\
  &\mbox{}\quad
  +\big[\tilde{M}_c^\mu- C^\mu+XG_0 (M_u^\mu +M_t^\mu +\tilde{M}^\mu_c )\big]_\Trans
  \label{eq:Mmu2}
\end{align}
where $\tilde{M}^\mu_c$ indicates the level of approximation of $M^\mu_c$.
Details for expanding $M^\mu_c$ in terms of various exchange-current
contributions are discussed in Ref.~\cite{hh2006}. The subtraction of $C^\mu$
in the transverse part $[\cdots]_\Trans$ avoids double-counting because, as
Eq.~(\ref{eq:Mmu1}) shows,  $C^\mu$ already provides a phenomenological account
for the mechanisms contained in $\tilde{M}^\mu_c$. This necessary corrective
term that follows from the exact equation (\ref{eq:MmuexactMC}) is not present
in any of the previous approximate treatments of gauge invariance. Its
importance is seen from the following considerations. The on-shell matrix
element for this approximation, for both real and virtual photons, can be
written simply as
\begin{align}
  \epsilon_\mu M^\mu &= \epsilon_\mu\big[ M^\mu_s +
  M_u^\mu+ M_t^\mu+ \tilde{M}^\mu_c
  \nonumber\\
  &\qquad\qquad\mbox{}+
   XG_0 \left(M_u^\mu+ M_t^\mu+ \tilde{M}^\mu_c \right)\big]_\Trans~,
   \label{eq:Mmu2real}
\end{align}
as can be seen from (\ref{eq:Transpol}), and thus becomes independent of
$C^\mu$. Without the subtraction, there would be an additional current $C^\mu$
in the transverse bracket $[\cdots]_\Trans$ of (\ref{eq:Mmu2real}) leading to
an unwarranted double counting of contributions.  Structurally,
Eq.~(\ref{eq:Mmu2real}) is similar to Eq.~(\ref{eq:Mmu1}) with $C^\mu\to
\tilde{M}^\mu_c$. In other words, the focus is shifted from constructing a
phenomenological GIP current $C^\mu$ to finding an acceptable approximation for
the exchange current contributions contained in $M^\mu_c$. Even though $C^\mu$
does not appear explicitly in (\ref{eq:Mmu2real}), this on-shell matrix element
may still be considered fully compliant with local gauge invariance because of
the underlying construction procedure in terms of (\ref{eq:Mmu2}). This is
particularly relevant if $M^\mu$ in the form of Eq.~(\ref{eq:Mmu2}) serves as
(off-shell) input for other reactions like, for example, two-meson production
processes~\cite{hh2019}, as explained in the subsequent discussion.

\section{Discussion}\label{sec:Discussion}

In summary, we have presented here an equivalent reformulation of the exact
expressions for the production current $M^\mu$ that cleanly separates, in a
model-independent manner, manifestly gauge-invariance-preserving tree-level
expressions from transverse final-state interaction (FSI) terms, with
microscopic reaction dynamics that are consistent for both real and virtual
photons. Since the technically challenging parts of the dynamics --- exchange
currents and FSI in a coupled-channel framework --- that generally require
approximations appear here only in the transverse part, approximations will
have no bearing on gauge invariance, for any level of sophistication of such
approximations.

This separation of tree-level GIP terms and transverse remainder may look
deceptively similar to what was suggested already in Ref.~\cite{hh2006},
however, there it was derived as an integral part of an elaborate approximation
scheme. The present findings show now that this splitting is an \emph{exact}
consequence of the fact that local gauge invariance is equivalent to an implied
(model-independent) longitudinal contact current $C^\mu_\Long$ given by
Eq.~(\ref{eq:CL2}). Hence, any phenomenological GIP current $C^\mu$ that
satisfies the necessary condition (\ref{eq:kCmu}) will provide a manifestly
gauge-invariant amplitude provided one subtracts its transverse part as in the
$[\cdots]_\Trans$ bracket of Eq.~(\ref{eq:MmuexactMC}) to make it
model-independent and avoid double counting when considering approximations of
$M^\mu_c$. This necessary subtraction is a new finding not present in previous
approximate treatments. As a consequence, the dependence on a GIP current
$C^\mu$  will only become manifest for approximations of the kind discussed in
the context of Eqs.~(\ref{eq:Mmu0}) and (\ref{eq:Mmu1}). For more sophisticated
approximations like in Eq.~(\ref{eq:Mmu2}) that explicitly take into acount
final-state interactions and exchange currents, the model-independent
logitudinal part of $C^\mu$ is implied but it does not appear explicitly in the
physical matrix element (\ref{eq:Mmu2real}).

However, if one employs the single-meson production current $M^\mu$ as a
contributing \emph{off-shell} mechanism for two-meson production~\cite{hh2019},
one must necessarily use the (off-shell) form (\ref{eq:Mmu2}) thus making the
dependence on $C^\mu$ manifest as an internal current mechanism. This will
ensure the overall \emph{local} gauge invariance of electromagnetic two-meson
production because, as shown in Ref.~\cite{hh2019}, the proof of the latter
necessarily relies on the validity of the gWTI~(\ref{eq:gWTI}) for all
contributing single-meson production currents.

More generally, off-shell single-meson production currents $M^\mu$ and their
contributing mechanisms are required as input for many reactions. Examples
including strangeness production and $NN$ bremsstrahlung processes are given in
Refs.~\cite{Nakayama2006a,Nakayama2009,Nakayama2011,Johansson2011,hh2012}.
Rendering each one of these processes gauge invariant in a consistent manner
necessarily requires that contributing single-meson production currents
reproduce the gWTI (\ref{eq:gWTI}) irrespective of the level of sophistication
that goes into determining $M^\mu$ in practice.

Maintaining local gauge invariance is an important and indispensable ingredient
for the consistent description of all electromagnetic processes. As we have
shown here, doing so for the single-meson production current $M^\mu$ is a
straightforward matter, in particular, when written in the form of
Eq.~(\ref{eq:MmuexactMC}) since the GIP part involves only the tree-level
diagrams of Fig.~\ref{fig:Mgip}.  In practice, the real issue is finding useful
approximations of all complex reaction mechanisms beyond the tree level to
provide meaningful results with reasonable effort.


\acknowledgments

The author gratefully acknowledges discussions with
Kanzo Nakayama (UGA). The present work was supported in part by the U.S.
Department of Energy, Office of Science, Office of Nuclear Physics, under Award
No.\ DE-SC0016582.

\bibliography{GenGauge}

\end{document}